\documentclass[final]{ws-ijmpa}
\usepackage[super,compress]{cite}
\usepackage{graphicx}
\usepackage{hyperref}
\usepackage{color}
\definecolor{IITred}{rgb}{0.5,0.05,0.05}
\definecolor{IITblue}{rgb}{0.05,0.05,0.4}
\newcommand{\alphas}{\ensuremath{\alpha_{\mathrm{s}}}}
\newcommand{\tev}{\ensuremath{\hbox{ TeV}}}
\newcommand{\gev}{\ensuremath{\hbox{ GeV}}}
\newcommand{\mev}{\ensuremath{\hbox{ MeV}}}
\def\iab{ab$^{-1}$}

\def\lum{cm$^{-2}$s$^{-1}$}
\newcommand{\fb}{\ensuremath{\hbox{ fb}}}
\newcommand{\smgg}{\ensuremath{\mathrm{SU(3)_c} \otimes \mathrm{SU(2)_L} \otimes \mathrm{U(1)}_Y}}
\newcommand{\cgg}{\ensuremath{\mathrm{SU(3)_c}}}

\newcommand{\ewgg}{\ensuremath{\mathrm{SU(2)_L} \otimes \mathrm{U(1)}_Y}}
\newcommand{\wigg}{\ensuremath{\mathrm{SU(2)_L}}}
\newcommand{\ygg}{\ensuremath{\mathrm{U(1)}_Y}}
\newcommand{\emgg}{\ensuremath{\mathrm{U(1)_{EM}}}}
\newcommand{\chisym}{\ensuremath{\mathrm{SU(2)_L\otimes SU(2)_R}}}
\newcommand{\gf}{\ensuremath{G_\mathrm{F}}}

\begin{document}
\markboth{Chris Quigg}{Scientific Overview}

%
\catchline{}{}{}{}{}
%

\title{Future Colliders Symposium in Hong Kong: Scientific Overview}

\author{Chris Quigg}

\address{Theoretical Physics Department, Fermi National Accelerator Laboratory\\ P.O. Box 500, 
Batavia, Illinois 60510 USA\footnote{quigg@fnal.gov \hfill \textsf{FERMILAB-CONF-16-033-T}}}

\maketitle


\begin{abstract}
Opening Lecture at the Hong Kong University of Science and Technology Jockey Club Institute for Advanced Study Program on High Energy Physics Conference, January 18--21, 2016.

\end{abstract}

\ccode{PACS numbers: 14.80.Bn, 29.20.db, 12.15.-y, 12.38.-t, 12.60.-i}

\section{Introductory Remarks}
I am grateful for the chance to open this conference on future colliders, especially because we will have a rich and stimulating program of talks on accelerator science and technology, on experimental results, plans, and detector concepts, and on the implications of  current theoretical understanding for the next decades of research.
\subsection{The Most Important Question}
One of my goals in this talk is to offer many questions, so I would like to begin with a big one that applies to everything we do---to theory, experiment, and accelerators alike---and to which we should give our scrupulous attention:

\vspace*{6pt}
\centerline{{\textit{How are we prisoners of conventional thinking?}}}
\vspace*{6pt}\noindent One aspect of this question pertains to the way we address well-identified problems. Refining an approach we have taken before may not be the optimal response to  known challenges. Just over the past few days I have been encouraged to hear mind-expanding ways of thinking about potential remedies for the high synchrotron-radiation-induced heat load in future proton--proton colliders, or of integrating final-focus beam elements with detectors in future electron--proton colliders. A second aspect has to do with the specific questions we are asking. Are we asking the right questions, or are we missing something essential? Have we framed our questions in the right way, or are we merely rehearsing conventional formulations, without re\"{e}xamining our premises and preconceptions?

\subsection{Accelerator Milestones \ldots}
Since we have come to Hong Kong to discuss future colliders, it is worth taking a moment to review the inventions, insights, and technologies that make this discussion possible. I heartily commend to your attention the volume by Sessler and Wilson~\cite{Sessler} for an authoritative, approachable, and considerably more complete survey.

The idea of cyclic acceleration embodied in Lawrence's cyclotron---more generally, that repeated applications of achievable gradients could accelerate charged particles to extremely high energies---underlies both the circular and linear colliders we are contemplating. 

A practical limitation of the cyclotron was the need to evacuate an entire cylindrical volume to accommodate the accelerating particle as it spiraled out from an initial small radius to a final large radius. By raising the confining magnetic field in synchrony with the increasing momentum, one could contain the particles in a beam pipe of fixed radius, dispensing with the hole in the doughnut. Coupled with the notion of phase stability---that particles lagging or leading the nominal phase (have less or more than the nominal energy) are accelerated more or less than the particles at nominal energy, the varying magnetic field leads to the idea of a synchrotron, the basis for all circular colliders. 

Early proton synchrotrons, such as the Berkeley Bevatron, still required apertures that were, by today's standards, gigantic. That changed dramatically with the invention of alternate-gradient (strong) focusing.\footnote{The classic description of how using quadrupole lenses to squeeze the beams sequentially in the horizontal and vertical planes leads to net focusing is in Ref.~\citen{Courant:1997rq}.} This advance put accelerator builders on the path to  dense, well-controlled  beams, making possible vacuum chambers only a few centimeters across. The subsequent development of active optics, including the breakthrough of stochastic cooling~\cite{vdM}, led to the intense beams required for high-luminosity colliders.

We take for granted the elementary fact that the c.m.\ energy of a beam of momentum $p$ incident on a fixed target of mass $M$ is $\sqrt{s_{\mathrm{ft}}} \approx \sqrt{2Mp}$, whereas the c.m.\ energy of beams in head-on collision is $\sqrt{s_{\mathrm{cb}}} \approx 2p$. Rolf Wider\"{o}e filed a patent application based on this observation in 1943.~\cite{Waloschek:1994qp} The concept of colliding beams was first realized at Frascati, Novosibirsk, and Stanford in the early 1960s.~\cite{Richter:1992eb,Shiltsev:2013vsa} Of these, the Princeton--Stanford Colliding Beams Experiment (CBX), carried out by G.~K.~O'Neill, C.~Barber, B.~Gittelman, and B.~Richter, entered my student consciousness through a story in the New York \textit{Times} reporting the first electron--electron collisions at a c.m.\ energy of $600\mev$.~\cite{CBXNYT} If that seems puny, consider that to achieve the same result in a fixed-target setting would require an electron beam of about $350\gev$, which we have still not attained! The \textit{Times} reported that ``two electrons come close enough for a collision only once every fifteen or twenty minutes." Soon thereafter, I read that the scientists did not know what would happen when they made the high-energy electrons collide head-on. ``What,'' I thought, ``could be more exciting than not knowing the answer?'' So began my fascination with high-energy colliders. Let us remember, when we seek to motivate new colliders, the power of {\textit{We do not know \ldots}}

To implement the Big Idea of particle colliders, we have required efficient radio-frequency accelerating cavities and superb vacuum technology, superconducting magnets and materials, and cryogenic technology. We have gone beyond the readily available stable beam particles, electrons and protons, using to excellent effect positrons and antiprotons as well. Perhaps we will see dedicated $\gamma\gamma$ colliders, muon storage rings as neutrino sources, and even $\mu^+\mu^-$ colliders. Novel acceleration methods may someday take us more efficiently to energies and luminosities of interest.

\subsection{Our Science Holds Many Opportunities}
About a decade ago, I was asked to present the issues before us to a panel charting the course for particle physics as part of the \textit{Physics 2010} decadal survey in the United States~\cite{EPP2010}. To illustrate the richness, diversity, and intellectual depth of our field,  the liveliness of our conversations with nearby disciplines, and the timeliness of our aspirations, I composed the list of goals shown as Figure~\ref{fig:wishlist}. 

\begin{figure}[b!]
\textsf{
\centerline{{\begin{minipage}[t]{0.489\textwidth}\noindent
\textcolor{IITblue}{In a decade or two, we can hope to \ldots}\\[-6pt]
\textcolor{IITred}{{\footnotesize{\phantom{M}\\
Understand electroweak symmetry breaking}\\		
		\textit{Observe the Higgs boson} \\		
		{Measure neutrino masses and mixings}\\		
		\textit{Establish Majorana neutrinos ($\beta\beta_{0\nu}$)}\\		
		{Thoroughly explore \textsf{CP} violation in $B$ decays} \\		
		\textit{Exploit rare decays ($K$, $D$, \ldots)}\\		
		{Observe $n$ EDM, pursue $e$ EDM}\\		
		\textit{Use top as a tool}\\		
		{Observe new phases of matter}\\		
		\textit{Understand hadron structure quantitatively}\\		
		{Uncover QCD's full implications}\\		
		\textit{Observe proton decay} \\		
		{Understand the baryon excess}\\		
		\textit{Catalogue matter and energy of universe}\\		
		{Measure dark-energy equation of state}\\		
		\textit{Search for new macroscopic forces}  \\		
		{Determine the (grand) unifying symmetry}}\\
		\phantom{M}}\\[-6pt]
		\textcolor{IITblue}{\ldots learn the right questions to ask} 
\end{minipage}%
\hfill
\begin{minipage}[t]{0.49\textwidth}\noindent
\textcolor{IITred}{{\footnotesize		\phantom{M}\\[-6pt]
\phantom{M}\\
	\textit{Detect neutrinos from the universe}\\	
	{Learn how to quantize gravity}\\	
		\textit{Learn why empty space is nearly weightless} \\		
		{Test the inflation hypothesis} \\		
		\textit{Understand discrete symmetry violation}\\		
		{Resolve the hierarchy problem}\\		
		\textit{Discover new gauge forces}\\
		{Directly detect dark-matter particles} \\		
		\textit{Explore extra spatial dimensions} \\		
		{Understand origin of large-scale structure} \\		
		\textit{Observe gravitational radiation} \\		
		{Solve the strong \textsf{CP} problem} \\		
		\textit{Learn whether supersymmetry is TeV-scale}\\		
		{Seek TeV-scale dynamical symmetry breaking}\\		
		\textit{Search for new strong dynamics}\\		
		{Explain the highest-energy cosmic rays}\\		
		\textit{Formulate the problem of identity}}}\\		
		\textcolor{IITblue}{{\hfill \ldots \hfill}\\[-6pt]
		\phantom{M}\quad\ldots\ and rewrite the textbooks!}
\end{minipage}}}}
\vspace*{1pt}
\caption{A to-do (wish) list for particle physics and neighboring fields, circa 2005. \label{fig:wishlist}}
\end{figure}
 Beyond the significance of individual entries, what is striking is the scale diversity and variety of experimental techniques, and the range of energies and distance scales involved. 

I hope you will agree that it is an impressive list of opportunities, including many for the LHC, and that it was plausible to anticipate very significant achievements over a twenty-year time horizon. Indeed, if we look back over the decade past, we and our scientific neighbors can claim a lot of progress. (I invite you to make your own report card!) Happily, and as expected, there is still much to accomplish. Please think about how you would update or improve the list, and how we can best advance the science.

\section{Discovery of the Higgs Boson in LHC Run 1}
We entered the LHC era having established two new laws of Nature, quantum chromodynamics and the electroweak theory. We had identified six flavors of quarks ($u, d, s, c, b, t$) and six flavors of leptons ($e, \mu, \tau$ and three neutrinos) as spin-$\frac{1}{2}$ fermions that we may idealize, provisionally, as pointlike particles. Interactions are derived from \smgg\ gauge symmetry, which reflects the curious fact that---in our experience---charged-current weak interactions apply only to the left-handed quarks and leptons. We do not know whether that reflects a fundamental asymmetry in the laws of Nature, or arises because right-handed charged-current interactions are so feeble that they have eluded detection.

The \cgg\ color symmetry that generates the strong interaction is unbroken, but the electroweak symmetry must be hidden because the weak interactions are short-range and standard Dirac masses for the quarks and leptons would conflict with the gauge symmetry. The surviving symmetry is the phase symmetry that generates electromagnetism: $\ewgg \to \emgg$. An essential task for the LHC has been to illuminate the nature of the previously unknown agent that hides electroweak symmetry. We have imagined a number of possibilities, including (i) A force of a new character, based on interactions of an elementary scalar; (ii) A new gauge force, perhaps acting on hitherto undiscovered constituents; (iii) A residual force that emerges from strong dynamics among electroweak gauge bosons; (iv) An echo of extra spacetime dimensions. The default option has been the first, an example of spontaneous symmetry breaking\footnote{See Ref.~\citen{Quigg:2015cfa} for a narrative of the historical development and references to the original literature.} analogous to the Ginzburg--Landau~\cite{GL} phenomenology of the superconducting phase transition and the Meissner effect.

\subsection{The Importance of the 1-Te$\!$V Scale}
The footprint of spontaneous symmetry breaking in the electroweak theory is the massive scalar particle known as the Higgs boson. While the electroweak theory does not predict the Higgs-boson mass, a thought experiment yields a conditional upper bound, or tipping point, for $M_H$.~\cite{Lee:1977eg} It is informative to consider scattering of longitudinal gauge bosons and Higgs bosons at high energies. The two-body reactions involving $W_{\mathrm{L}}^+W_{\mathrm{L}}^-, Z_{\mathrm{L}}Z_{\mathrm{L}}, HH, HZ_{\mathrm{L}}$ satisfy s-wave unitarity, provided that $M_H \le \left(8\pi\sqrt{2}/3\gf\right)^{1/2} \approx 1\tev$.  If the bound is respected, perturbation theory is reliable (except near resonance poles), and a Higgs boson is to be found below $1\tev$ in mass. If not, weak interactions among $W_{\mathrm{L}}^\pm, Z_{\mathrm{L}}, H$ become strong on 1-TeV scale. One way or the other, \textit{new phenomena are to be found around} $1\tev$. This analysis shows us that the role of the ``Brout--Englert--Higgs mechanism'' in the electroweak theory is not only to break $\ewgg \to \emgg$ and to generate masses for the electroweak gauge bosons and the fermions, but also---through the action of the Higgs boson---to regulate gauge boson interactions at high energies. 
 
 In the years leading up to experiments at the LHC, the analysis of precise measurements of electroweak observables, within the standard electroweak theory, pointed to a light Higgs boson, with a mass no greater than about $200\gev$.\footnote{See, for example, Refs.~\citen {Group:2008aa} and \citen{Flacher:2008zq}.}

We have not (yet) found an argument---based either on theoretical consistency or on the analysis of observations within a particular framework---that points to a specific scale beyond the 1-TeV scale.
\subsection{Searches at the Large Hadron Collider}
Let us quickly review what experiments at the LHC have revealed so far about the Higgs boson. The LHC makes possible searches in many channels of production (gluon fusion $gg \to H$, associated production $q\bar{q}^\prime \to H(W,Z)$, vector-boson fusion, and the $Ht\bar{t}$ reaction) and decay ($\gamma\gamma, WW^*, ZZ^*, b\bar{b}, \tau^+\tau^-$, \ldots). Since the discovery of $H(125) \to (\gamma\gamma,\ell^+\ell^-\ell^+\ell^-)$ was announced by the ATLAS~\cite{Aad:2012tfa} and CMS~\cite{Chatrchyan:2012xdj} Collaborations in 2012, the evidence has developed as it would for a standard-model Higgs boson.~\cite{ATLASH,CMSH}

In addition to the $\gamma\gamma$ and $ZZ$ discovery modes, the $W^+W^-$ mode~\cite{Chatrchyan:2013iaa,ATLAS:2014aga} is established, and the spin-parity assignment  $J^P = 0^+$ is overwhelmingly favored.~\cite{Aad:2015mxa,Khachatryan:2014kca} A combined measurement of the Higgs-boson mass yields $M_H = 125.09 \pm 0.24\gev$.~\cite{Aad:2015zhl}  A grand average of the combined ATLAS and CMS measurements of the Higgs-boson signal yield (i.e., production times branching fraction) is $1.09 \pm 0.11$ times the standard-model expectation.~\cite{mucombo} Within the uncertainties, individual modes are in line with the standard-model predictions.

If $H(125)$ is to be unambiguously identified as the standard-model Higgs boson of our textbooks, what remains to be demonstrated? We need to investigate, through precise measurements of the $HWW$ and $HZZ$ couplings, whether it fully accounts for electroweak symmetry breaking. We must extend the indications~\cite{mucombo} that $H(125)$ couples to fermions, test whether the $Hf\bar{f}$ couplings are proportional to the fermion masses, and indeed whether the interaction of fermions with the Higgs field accounts entirely for their masses. The predicted branching fractions are collected in Table~\ref{tab:HBR}.
\begin{table}[tb]
\tbl{Branching fractions $\mathcal{B}$ for a 125-GeV standard-model Higgs boson (from Ref.~\citen{Dittmaier:2012vm}).  
\label{tab:HBR}}
 {\begin{tabular}{lccccccccc}
 \toprule
Mode & $b\bar{b}$ & $WW$  & $gg$ & $\tau^+\tau^-$ & $c\bar{c}$ & $ZZ$ &  $\gamma\gamma$ & $Z\gamma$  & 
 $\mu^+\mu^-$  \\
 \colrule
$\mathcal{B}$ &
$0.577$ & $0.215$ & $0.0857$ & $0.0632$ & $0.0291$ & $0.0264$ & $0.00228$ & $0.00154$ &  $0.00022$ \\
 \botrule
 \end{tabular}}
\end{table}
It is noteworthy, and completely expected at the current level of sensitivity, that we have only observed Higgs couplings to fermions of the third generation---top from the production rate attributed to gluon fusion, direct observations of decays into $b\bar{b}$ and $\tau^+\tau^-$. It is essential to learn whether the same mechanism is implicated in the masses of the lighter fermions. Detection of $H \to \mu^+\mu^-$ is foreseen at the LHC. The observation of the decay into charm pairs looks highly challenging in the LHC environment, but merits very close consideration.

Another significant test is the total width of $H(125)$, predicted to be $\Gamma(H(125))=4.07\mev$ for a standard-model Higgs boson. This is well below the experimental resolution for a direct determination at the LHC, but by applying the clever insight that---within a framework that resembles the standard model---measurements of the off-shell coupling strength in the $WW$ and $ZZ$ channels at invariant masses above $M_H$ constrain the Higgs-boson width, the LHC experiments restrict $\Gamma(H(125))$ to be less than a few tens of MeV.\footnote{For early theoretical analyses, see Ref.~\citen{Caola:2013yja,Kauer:2012hd}. First experimental determinations are presented in Refs.~\citen{Khachatryan:2014iha} and \citen{Aad:2015xua}.}

We will continue to search for admixtures of spin-parity states other than the dominant $J^P = 0^+$, and to test that all production modes are as expected.

Much exploration remains as well. Does $H(125)$ have partners? Does it decay to new particles, perhaps serving as a portal to unseen sectors? Are there any signs of compositeness, of new strong dynamics? Finally, we can contemplate the implications of a 125-GeV Higgs boson.

\subsection{Why does discovering the agent of electroweak symmetry breaking matter?}
An instructive way to respond to this question is to imagine a world without a symmetry-breaking (Higgs) mechanism at the electroweak scale. A full analysis of that \textit{Gedanken} world is rather involved,~\cite{Quigg:2009xr} but here are the main points, restricted for simplicity to one generation of quarks and leptons. The electron and quarks would have no mass. QCD would confine quarks into nucleons and other hadrons, and the nucleon mass---to which the up- and down-quark masses contribute only small amounts in the real world---would be little changed.\footnote{Whether the proton or neutron would be the lighter---hence stable---nucleon is too close a call for us to settle.} In the Lagrangian, the massless quarks exhibit an 
\chisym\ chiral symmetry that is spontaneously broken, near the confinement scale, to SU(2) isospin symmetry. The resultant linkage of left-handed and right-handed quarks gives rise to the ``constituent-quark'' masses, and hides the electroweak symmetry because the left-handed and right-handed quarks transform differently under \ewgg. The electroweak gauge bosons $W^\pm$ and $Z$ acquire tiny masses, about $2\,500$ times smaller than those we observe in the real world. The scale is set, not by the vacuum expectation value $v$ of the (absent) Higgs field, but by the pion decay constant $f_\pi$.
        
Now suppose that protonuclei---say, alpha particles---are created in the early universe and survive to late times (whatever that might mean). A massless electron means that the Bohr radius of an atom would be infinite, so it is not possible to identify an electron as belonging to a specific atom. In other words, ``atoms'' lose integrity. If an electron can't be assigned to a particular nucleus, the notion of valence bonding evaporates. No atoms means no chemistry, no stable composite structures like liquids, solids, \ldots 
no template for life! 

Returning to our world, it's important that we not get ahead of the evidence. We have good indications that $H(125)$ couples approximately as expected to top and bottom quarks and to the tau lepton. We anticipate that $H \to \mu^+\mu^-$ can be established at the High-Luminosity LHC, if not before. Measuring the coupling of $H(125)$ to charm seems highly challenging at the LHC; to achieve that, we need either new insights or a Higgs factory. Demonstrating $H \to e^+e^-$, with its predicted branching fraction $\approx 5 \times 10^{-9}$, is beyond challenging, but to my mind showing that spontaneous symmetry breaking gives mass to the electron would merit a Nobel Prize in Chemistry!
\section{Looking ahead to future colliders}

\subsection{A Higgs Factory?}
The discovery of $H(125)$ motivates consideration of an $e^+e^-$ Higgs factory (or a stage of a linear collider), and comparison with what LHC will do, and when experiments will happen. An excellent starting point is Ref.~\citen{Dawson:2013bba}. The initiatives under active discussion include the International Linear Collider in Japan,~\cite{ilc} the Circular Electron--Positron Collider in China,~\cite{cepc} and the FCC-ee Design Study centered at CERN.~\cite{fcc-ee} The performance of a Higgs factory is addressed in the following talk by Matt Reece, so my comments will be brief.

There is little question that if a Higgs factory allowing the detailed study of the associated-production reaction $e^+ e^- \to HZ$ at $\sqrt{s} \approx 240\gev$ were available today, it would be a superb complement to ongoing experiments at the LHC, and would attract many users. That is not the case, and so we need to assess what a purpose-built machine can do when its experimental program begins. At any moment, the telling measurements depend on what is already know. For example, will $H(125)$ continue to match the textbook description, or will it begin to show nonstandard properties?  How would the discovery of another ``Higgs-like object'' change the picture? And what would direct evidence for or against new degrees of freedom mean for our goals for a new machine?

It's also worthwhile to examine the benefits and (opportunity) cost of parameter variations for a projected collider. What would be the value of extending a Higgs factory to the top threshold? How well could we hope to determine the top mass, the strong coupling $\alphas$, and the top Yukawa coupling to the Higgs boson? How much running time would be required? If we imagine running on the $Z$-peak, what is the goal---Giga-$Z$ or Tera-$Z$? What are the implications of high-luminosity running at the $Z$ for the machine design? How much running time would be required to significantly improve our knowledge of electroweak observables, or to exploit the copious source of boosted $b$-hadrons? What would it take to significantly improve the precision of $M_W$ by mapping the excitation curve?

Many of these Higgs-factory enhancements are easy to dream about, but may be costly to deliver and take much time to exploit!
\subsection{A Very Large Hadron Collider: Generalities}

We are looking beyond CERN's Large Hadron Collider to a ring with circumference two to four times that of the LHC. Superconducting dipoles with field strength between 15 and 20 teslas would support a proton--proton collider with c.m.\ energy $\sqrt{s} \approx 50\hbox{ to }100\tev$.\footnote{Such dipoles could enable a 33-TeV $pp$ collider in the LHC tunnel. See Ref.~\citen{Gourlay} for a reality check.} The goal of a ``100-TeV'' hadron collider has been set for a machine study, but it important to keep in mind that a feasible, scientifically desirable $pp$ collider might have a different energy. At this point, we should undertake physics studies over a range of energies, bearing in mind that different combinations of energy and luminosity can, to some degree, yield comparable discovery potential.\footnote{In Ref.~\citen{Eichten:1984eu}, back at the beginning of time, we explored the reach of $p^\pm p$ colliders at $\sqrt{s} = 2, 10, 20, 40, 70, \hbox{and }100\tev$. I am pleased to note that the LHC Higgs Cross Section Working Group, Ref.~\citen{HXSWG}, is providing cross sections at $\sqrt{s} = 14, 33, 40, 60, 80, \hbox{and }100\tev$.} The work we do for ``100-TeV'' can enhance what we achieve with LHC. It is important to consider search and measurement examples that will stretch detector capabilities, to examine the role of special-purpose detectors, including concepts that have been set aside in the past.~\cite{FELIX} There is great value in developing tools that enable others to extend the work.

I think it premature to enunciate the scientific case for the ``100-TeV'' hadron collider, but the right time to explore possibilities. We can cite plenty of reasons that such a machine could be highly exciting and scientifically rewarding,\footnote{See Ref.~\citen{Arkani-Hamed:2015vfh} for a good start.} but it will be many years before we will be able to make a credible technical proposal. It is overwhelmingly likely that we will learn a great deal in the intervening time (not only from the LHC), and I would not bet against discoveries that alter our conception of the Great Questions in some dramatic way. What we learn from the LHC (and elsewhere) might point to an energy landmark for the next great machine. I recall that for nearly two decades, the central pillar of the case put forward for a linear collider has been that it would unravel the rich spectrum of light superpartners. That case has vanished identically. 

In a world with multiple, widely separated, physical scales, the electroweak scale and the light Higgs-boson mass present a puzzle: Why are $M_W$ and $M_H$ so much smaller than the unification scale or the Planck scale, presuming those to be physically significant? In quantum field theory, distant scales tend to be linked through quantum corrections, and so large hierarchies seem to ask for ``natural'' explanations\footnote{See Ref.~\citen{Dine:2015xga} for a perceptive review of the naturalness principle.} more satisfying than ``just-so stories.'' We haven't yet found any direct evidence on the 1-TeV scale for new dynamics or a new symmetry that could explain the many orders of magnitude between the electroweak scale and the others. (Supersymmetry, in particular, is hiding very effectively.) Experiment has not established a  pattern of serious quantitative failures of electroweak theory, nor have we uncovered any clear sign of the flavor-changing neutral currents that occur generically in ``new-physics'' extensions to the standard model. Searches for forbidden or suppressed processes that might reveal something about flavor-changing neutral currents are consequently of great interest, as is the ongoing campaign to make ever-more-precise tests of the electroweak theory.

Opinions about how to respond to a possible hierarchy problem have evolved over many years. We originally sought once-and-done remedies, such as supersymmetry or technicolor, that invoked new physics on the TeV scale to exorcise the problem once and for all. Maybe that is not the right approach. Should we instead favor a stepwise approach with a sequence of effective theories? Might we have misunderstood the hierarchy problem, and so need to reframe it? Perhaps it is time to ask whether the unreasonable effectiveness of the standard model~\cite{Quigg:2009vq} (to borrow a turn of phrase from Eugene Wigner~\cite{wigner}) is itself a deep clue to what lies beyond.

All this is to argue that we should continue to examine our notion of the hierarchy problem. It is, after all, a problem for our feelings about how nature should work, not a contradiction that we arrive at from first principles. Ken Wilson, one of the founders of naturalness, continued to think about the issues. For a revealing counterpoint to appeals to authority, see \S5 of his historical survey, Ref.~\citen{Wilson:2004de}, in the passage beginning with ``The final blunder \ldots\ .''

How we conceive of the hierarchy problem will help determine how theorists invest their intellectual capital. But I am skeptical of the assertion that---by some arbitrary measure---the 100-TeV machine will test naturalness at the $10^{-4}$ level, rather than 1\% at the LHC, and that those two orders of magnitude will somehow settle the matter. As a justification for a new collider, it is unpersuasive.

While the standard model gives an excellent account of a wealth of experimental information, it has nothing to say about a number of important questions, including the nature of dark matter, the origin of the matter excess in the universe, the riddle of dark energy, and the pattern of fermion masses and mixing angles. And although the LHC---still in its exploratory phase---hasn't yet presented us with new physics on the TeV scale, we may have some hints.~\cite{Koppen}

Both CMS~\cite{Khachatryan:2014gha} and ATLAS~\cite{Aad:2015owa} report indications of excesses in diboson invariant mass distributions in the neighborhood of $2\tev$ in their event samples at $\sqrt{s} = 8\tev$. The most recent data of the LHC$b$ Experiment~\cite{Aaij:2015bfa} display continuing tensions in the quark-mixing matrix element $V_{ub}$ measured by different techniques. The ratio of branching fractions $\mathcal{B}(B^{+}\rightarrow K^{+}\mu^{+}\mu^{-})/\mathcal{B}(B^{+}\rightarrow K^{+}e^{+}e^{-})$ in the interval $1\gev^2 \le q^2 \le 6\gev^2$ is determined by LHC$b$ as $0.745^{+0.090}_{-0.074}\mathrm{\,(stat)}\pm0.036\mathrm{\,(syst)}$, which differs by $2.6\sigma$ from the lepton-universality expectation.~\cite{Aaij:2014ora} Both ATLAS~\cite{ATLASdigamma} and CMS~\cite{CMS:2015dxe} have shown provocative indications of a diphoton resonance near $750\gev$ in their 2015 run at $\sqrt{s} = 13\tev$. If real, this will be a sensational discovery on its own, and will almost certainly indicate other new phenomena to follow.

\subsection{A Very Large Hadron Collider: Some Specifics}
\begin{figure}
\centerline{\includegraphics[width=0.8\textwidth]{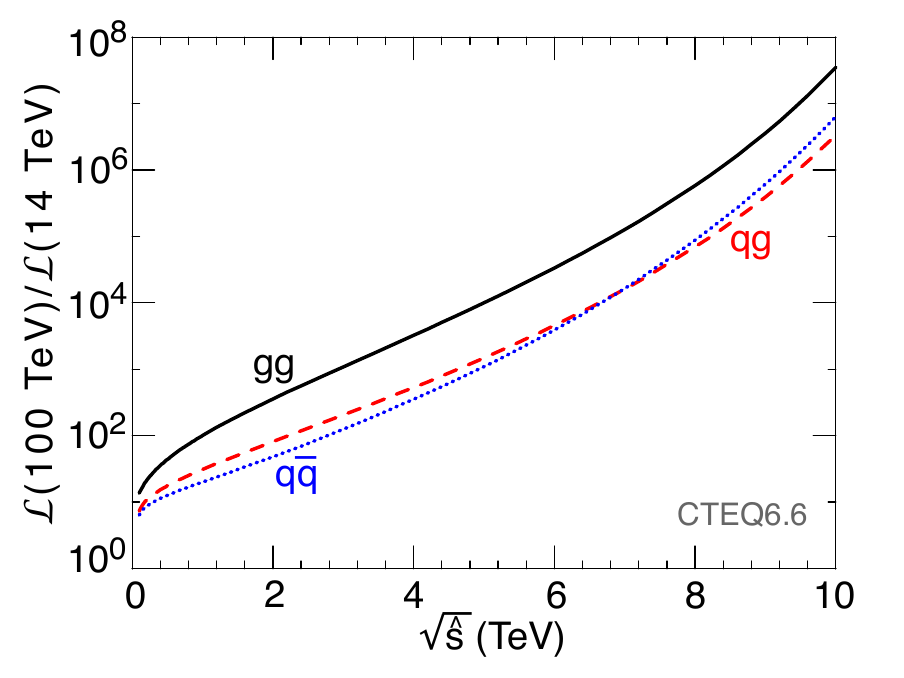}}
\caption{Parton luminosity ratios [from Ref.~\citen{Hinchliffe:2015qma}] at $\sqrt{s} = 100\hbox{ and }14\tev$ as a function of parton--parton subenergy $\sqrt{\hat{s}}$, evaluated using the CTEQ6.6 parton distributions~\cite{Nadolsky:2008zw} with $Q^2=\hat{s}$. \label{fig:lumrats}}
\end{figure}

What new opportunities will a ``100-TeV'' $pp$ collider offer? Figure~\ref{fig:lumrats} shows the ratios of parton luminosities for collisions of $gg$, $q\bar{q}$, and $qg$ (gluons $g$ and light quarks $q$) in $pp$ collisions at $\sqrt{s} = 100\hbox{ and }14\tev$.
At what will be modest parton subenergies at a 100-TeV collider, $\sqrt{\hat{s}} \lesssim 1\tev$, the parton luminosities increase by an order of magnitude or more. This advantage could, in principle, be overcome by increasing the 14-TeV $pp$ luminosity by 1--2 orders of magnitude beyond the High-Luminosity LHC, but that is a somewhat daunting prospect. At higher values of $\sqrt{\hat{s}}$, there is a decisive advantage to increasing
$\sqrt{s}$.

An instructive example at modest scales is the increase in Higgs-boson production cross sections shown in Table~\ref{tab:H}.\footnote{Michelangelo Mangano shows how we can expect to refine our knowledge of Higgs-boson properties in his talk at this symposium.} 
\begin{table}[tbh]
\tbl{Ratio $\mathcal{R}_{100:14} \equiv \sigma(\sqrt{s}=100\tev)/\sigma(\sqrt{s}=14\tev)$  for various Higgs-production reactions, according to Ref.~\citen{HXSWG}.
\label{tab:H}}
 {\begin{tabular}{lccccccc}
 \toprule
Process & $gg\to H$ & $q\bar{q}\to WH$ & $q\bar{q}\to ZH$  & $qq\to qqH$  &
 $t\bar{t}H$ & $b\bar{b}H$& $gg\to HH$ \\
 \colrule
$\mathcal{R}_{100:14}$ &
14.7 & 9.7 & 12.5 & 18.6 & 61 & 15 & 42 \\
 \botrule
 \end{tabular}}
\end{table}
Beyond giving us the means to learn more about $H(125)$ and other particle that come into view at the LHC, a 100-TeV--class collider will enhance the discovery reach at low masses, making accessible rare processes and phenomena characterized by low detection efficiencies and challenging backgrounds.

Consider as well particles in the upper reaches of the HL-LHC
discovery range, for example a gauge boson of mass around parton
subenergy $\sqrt{\hat{s}}=6\tev$ produced singly in the $q\bar{q}$
channel, or pair production of $\approx 3\tev$ particles in the $gg$
channel, for which the parton luminosities increase by factors of $10^4$ and $10^5$, respectively. If we contemplate an order-of-magnitude increase in the integrated $pp$ luminosity, this implies event samples up to a million times larger.

At still higher energy scales, the 100-TeV collider enters unexplored terrain, where we may find new particles and new phenomena.\footnote{Matthew McCullough exhibited a selection of these in his talk at the Symposium.} In addition to all the usual suspects of the LHC era---supersymmetry, strong dynamics, extra dimensions, and all the rest---we might have access to $(B+L)$-violating phenomena. Tye and Wong, for example, have argued that a 9.3-TeV sphaleron produced in collisions of left-handed light quarks would give rise to final states containing multiple same-sign leptons and multiple $b$ quarks.~\cite{Tye:2015tva}

New phenomena may arise with relatively large cross sections, should hitherto unknown collective effects emerge as increasing energies create unusual conditions in proton--proton collisions. I have in mind novel event structures, perhaps reflecting the partonic structure of the protons, or evidence for new component of particle production such as thermalization or hydrodynamical behavior.

\begin{figure}
\centerline{\includegraphics[width=0.5\textwidth]{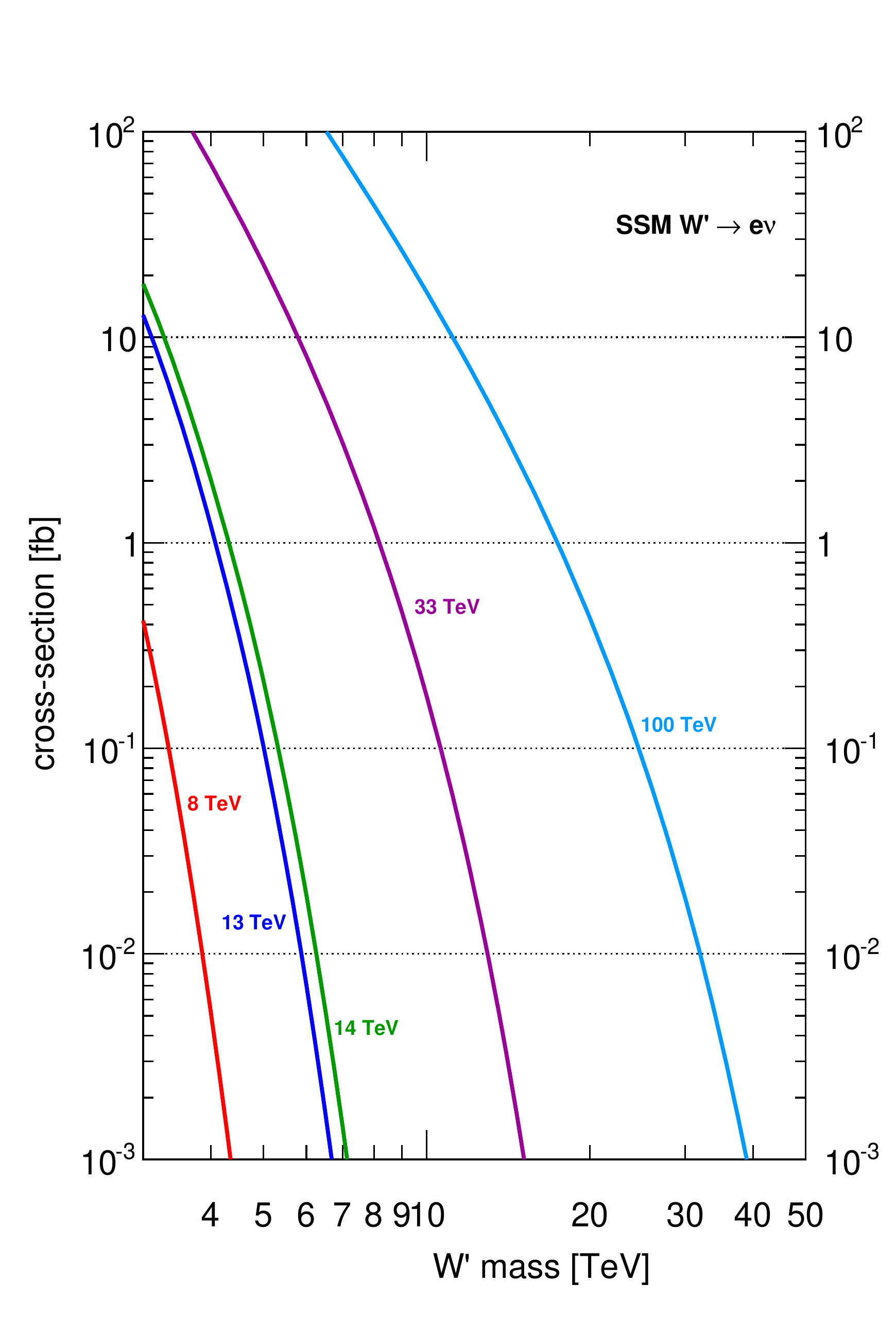}\hfil\includegraphics[width=0.50\textwidth]{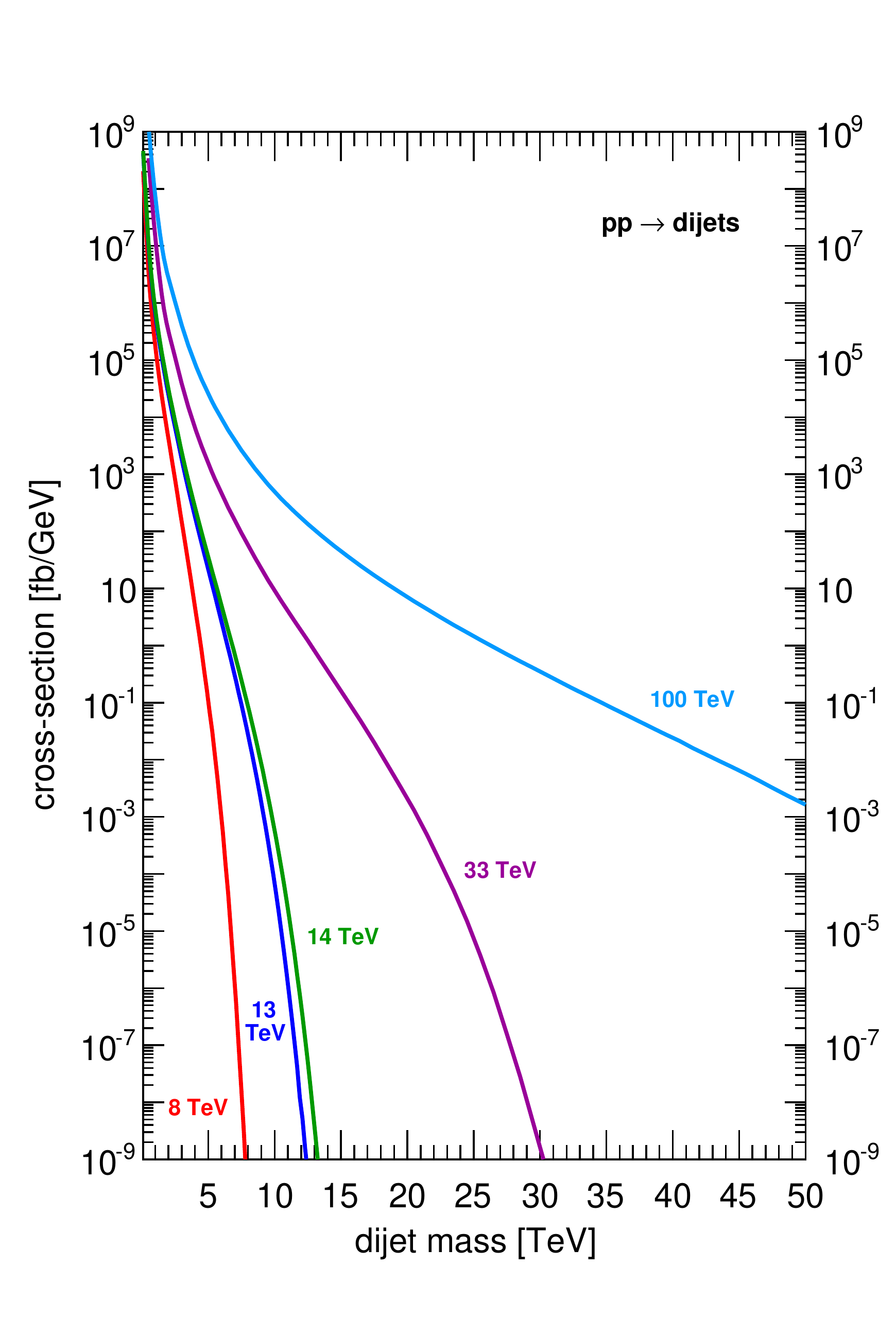}}
\caption{Left pane: Cross section for the production and decay of a sequential standard-model $W^\prime \to e\nu$ boson calculated at next-to-leading order with \textsf{MCFM}\cite{Campbell:2011bn}, at c.m.\ energies $\sqrt{s} = 8, 13, 14, 33, 100\tev$. Right pane: Cross section $d\sigma/d\mathcal{M}$ for the production of dijets with invariant mass $\mathcal{M}$ calculated at next-to-leading order with \textsf{MCFM}\cite{Campbell:2011bn}, at c.m.\ energies $\sqrt{s} = 8, 13, 14, 33, 100\tev$.\label{fig:wprime}}
\end{figure}

I show in Figure~\ref{fig:wprime} two examples of how the discovery reach increases as the 
$pp$ energy is raised beyond $\sqrt{s} = 14\tev$. The left pane depicts the cross section times branching fraction at next-to-leading order for a sequential standard-model $W^\prime$-boson decaying into electron $+$ antineutrino---an artificial benchmark, but one that is straightforward to state and adapt to other cases.\footnote{This stylized $W^\prime$ has standard-model couplings to fermions, but no decays into gauge bosons.} If the discovery limit at the 14-TeV HL-LHC is taken to be $7\tev$, then (at constant branching fraction and $pp$ luminosity) the 100-TeV limit would be approximately $30\tev$.
With an order of magnitude increase in integrated luminosity, the discovery limit approaches $40\tev$.

 The right pane of Figure~\ref{fig:wprime} shows the dijet invariant mass distribution evaluated at next-to-leading order. A 5-TeV reach in dijet mass at the HL-LHC  grows to $20\tev$ at $\sqrt{s} = 100\tev$, with fixed $pp$ luminosity, while $10\tev$ at the HL-LHC increases to over $50\tev$ at the 100-TeV collider, opening much space for discovery.

\subsection{Provisional Luminosity Recommendations}
At the 2015 Hong Kong workshop, we examined various arguments for the luminosity that would be required for a productive 100-TeV hadron collider. Our assessment,~\cite{Hinchliffe:2015qma} which should be revisited during the ongoing studies, was this:
\begin{quote}
``The goal of an integrated luminosity in the range of 10-20~\iab\ per
experiment, corresponding to an ultimate instantaneous luminosity~\cite{benedikt}
approaching $2\times 10^{35}$~\lum\ seems well-matched
to our current perspective on extending the discovery reach for new
phenomena at high mass scales, high-statistics studies of possible new
physics to be discovered at (HL)-LHC, and incisive studies of the
Higgs boson's properties.  Specific measurements may set more
aggressive luminosity goals, but we have not found generic arguments
to justify them. The needs of precision physics arising from new
physics scenarios to be discovered at the HL-LHC, to be suggested by
anomalies observed during the $e^+e^-$ phase of a future circular
collider, or to be discovered at 100~TeV, may well drive the need for
even higher statistics. Such requirements will need to be established
on a case-by-case basis, and no general scaling law gives a robust
extrapolation from 14 TeV. Further work on \textit{ad hoc} scenarios,
particularly for low-mass phenomena and elusive signatures, is
therefore desirable.''
\end{quote}
\subsection{Hadron Colliders and Unified Theories}
The neutrality of matter---with its implication that proton and electron charges exactly balance---is a powerful encouragement for a unified theory of the strong, weak, and electromagnetic interactions. An attractive possibility is a simple unifying gauge group $\mathcal{G}$ that contains the \smgg\ bits that we have discovered in our relatively low-energy experiments. Taking into account the degrees of freedom we know, the appropriately normalized \cgg, \wigg, and \ygg\ coupling constants evolve toward a common value at very high energies. Coupling-constant unification is more promising in supersymmetric $\mathrm{SU(5)}$ than in the original $\mathrm{SU(5)}$ theory, provided that the change in evolution due to a full spectrum of superpartners occurs near $1\tev$.~\cite{Raby:2009sf}

Plotted as a function of $\ln{Q}$,  $1/\alphas$ evolves with slope $7/2\pi$ if the standard model is embedded in $\mathrm{SU(5)}$, but the slope changes to $3/2\pi$ above the energy at which a full spectrum of superpartners is active. Could experiments at the LHC, or a future collider, test the hypothesis of supersymmetric unification by measuring the strong coupling constant (or the weak mixing parameter $\sin^2\theta_{\mathrm{W}}$) as a function of scale? ATLAS~\cite{ATLAS:2013lla} and CMS~\cite{Khachatryan:2014waa} have already made what I would characterize as exploratory measurements of $\alphas$ that extend to scales above $1\tev$ by determining the ratio of three-jet to two-jet rates in $pp$ collisions at $\sqrt{s}=7\tev$.

Seeing, or not seeing, a change of slope would be powerful evidence for or against the existence of a new set of colored particles that would complement ongoing searches for specific new-particle signatures. Considerable thought will be required to determine the most promising classes of measurements. I suspect that the study of $Z^0 + \hbox{jets}$ will be fruitful. A continuing conversation between theory and experiment will be needed to isolate $\alphas(Q)$ measured at a high scale.
\section{Issues for the Future (Starting Now!)}
Let us conclude with a short list of questions we would like to answer:
\begin{enumerate}
\item  \emph{There is a Higgs boson!} Might there be several? 
\item    Does the Higgs boson regulate $WW$ scattering at high energies?
\item   Is the Higgs boson elementary or composite? How does it interact with itself? What triggers electroweak symmetry breaking?
\item   Does the Higgs boson give mass to fermions, or only to the weak bosons? What sets the masses and mixings of the quarks and leptons? (How) is fermion mass related to the electroweak scale?
\item   Will new flavor symmetries  give insights into fermion masses and mixings?
\item   What stabilizes the Higgs-boson mass below 1 TeV?
\item   Do the different charged-current behaviors of left-handed and right-handed fermions reflect a fundamental asymmetry in Nature's laws?
\item    What will be the next symmetry that we recognize? Are there additional heavy gauge bosons? Is nature supersymmetric? Is the electroweak theory contained in a unified theory of the strong, weak, and electromagnetic interactions?
 \item   Are all flavor-changing interactions governed by the standard-model Yukawa couplings? Does ``minimal flavor violation'' hold? If so, why? At what scale?
 \item   Are there additional sequential quark and lepton generations? Or new exotic (vector-like) fermions?
\item    What resolves the strong \textsf{CP} problem?
\item    What are the dark matters? Is there any flavor structure?
\item     Is electroweak symmetry breaking an emergent phenomenon connected with strong dynamics? How would that alter our conception of unified theories of the strong, weak, and electromagnetic interactions?
\item     Is  electroweak symmetry breaking related to gravity through extra spacetime dimensions?
\item     What resolves the vacuum energy problem?
\item     (When we understand the origin of  electroweak symmetry breaking,) what lessons does  electroweak symmetry breaking hold for unified theories? \ldots\ for inflation? \ldots\ for dark energy?
\item     What explains the baryon asymmetry of the universe? Are there new (charged-current) \textsf{CP}-violating phases?
 \item     Are there new flavor-preserving phases? What would observation, or more stringent limits, on electric-dipole moments imply for theories beyond the standard model?
\item      (How) are quark-flavor dynamics and lepton-flavor dynamics related (beyond the gauge interactions)? 
\item      At what scale are  neutrino masses set? Do they speak to the TeV scale, the unification scale, the Planck scale, or \ldots?
 \item     Could our Laws of Nature be environmentally determined?\\[6pt]
 And finally, the question that looms over all the others, \\
\centerline{\emph{How are we prisoners of conventional thinking?}}
\end{enumerate}

\section*{Acknowledgments}

Fermilab is operated by Fermi Research Alliance, LLC under Contract No. De-AC02-07CH11359 with the United States Department of Energy. I thank the conference organizers for the kind invitation to speak. I am grateful to Henry Tye and members of the Jockey Club Institute for Advanced Study for their generous hospitality, to the participants for their contributions to a stimulating environment, and to Prudence Wong for her gracious practical assistance. I thank John Campbell for providing Figure~\ref{fig:wprime}, and for helpful discussions.

\appendix



\begin{thebibliography}{00}  

\bibitem{Sessler} A.~Sessler and E.~Wilson, \textit{Engines of Discovery: A Century of Particle Accelerators} (World Scientific, Singapore, 2007).
  
\bibitem{Courant:1997rq} 
  E.~D.~Courant and H.~S.~Snyder,
  ``Theory of the alternating gradient synchrotron,''
  Annals Phys.\  {\bf 3}, 1 (1958)
  [Annals Phys.\  {\bf 281}, 360 (2000)] \url{http://j.mp/1U6Y3ON}.

\bibitem{vdM} S.~van~der~Meer, ``Nobel Lecture: Stochastic Cooling and the Accumulation of Antiprotons,''  \url{http://j.mp/1VHCbYF}.

\bibitem{Waloschek:1994qp} 
  P.~Waloschek (ed.) and R.~Wider\"{o}e,
  ``The Infancy of particle accelerators: Life and work of Rolf Wider\"{o}e,'' DESY-94-039.
  The Colliding Beams patent, \textit{Anordnung zur Herbeif\"{u}hrung Kernreaktionen,}
 submitted 6 September 1943 and issued 11 May 1953, can be seen as pp.~95 and 96 of \url{http://j.mp/1Q0EkPg}.

\bibitem{Richter:1992eb} 
  B.~Richter,
  ``The Rise of Colliding Beams,''
  in \textit{The Rise of the Standard Model: Particle physics in the 1960s and 1970s,} eds.  L.~H.~Hoddeson, L.~Brown, M.~Riordan and M.~Dresden (Cambridge University Press, Cambridge, 1997) pp.~261--284.
   
\bibitem{Shiltsev:2013vsa} 
  V.~Shiltsev,
  ``The first colliders: AdA, VEP-1 and Princeton-Stanford,'' to appear in \textit{Challenges and Goals for Accelerators in the XXI Century,} Eds. O.~Br\"{u}ning and S.~Myers (World Scientific, Singapore, 2016)  [arXiv:1307.3116 [physics.hist-ph]].
  
\bibitem{CBXNYT} ``Atom Smasher Test Shows Way to Save on Energy,'' New York  \textit{Times,} March 13, 1965, p.~9.



\bibitem{EPP2010} \textit{EPP 2010: Elementary Particle Physics in the 21st Century} (National Research Council, Washington, 2006) \url{http://sites.nationalacademies.org/BPA/BPA_048230}.  


\bibitem{Quigg:2015cfa} 
C.~Quigg,
  ``Electroweak Symmetry Breaking in Historical Perspective,''
  \textit{Ann.\ Rev.\ Nucl.\ Part.\ Sci.}  {\bf 65}, 25 (2015)
  [arXiv:1503.01756 [hep-ph]].


\bibitem{GL} V.~L.~Ginzburg and L.~D.~Landau, \emph{Zh.
Eksp.  Teor.  Fiz.} \textbf{20}, 1064 (1950); English translation: see
\emph{Men of Physics: Landau,} Vol.  II, edited by D. ter Haar,
(Pergamon, New York, 1965).

\bibitem{Lee:1977eg} 
  B.~W.~Lee, C.~Quigg and H.~B.~Thacker,
  ``Weak Interactions at Very High Energies: The Role of the Higgs Boson Mass,''
  \textit{Phys.\ Rev.\ D}{\bf 16}, 1519 (1977).

\bibitem{Group:2008aa} 
  ALEPH and CDF and D0 and DELPHI and L3 and OPAL and SLD and LEP Electroweak Working Group and Tevatron Electroweak Working Group and SLD Electroweak Working Group and Heavy Flavour Group Collaborations],
 ``Precision Electroweak Measurements and Constraints on the Standard Model,''
  arXiv:0811.4682 [hep-ex]. See also \url{http://lepewwg.web.cern.ch} and \url{http://sanc.jinr.ru/users/zfitter/}.
  

\bibitem{Flacher:2008zq} 
  H.~Fl\"{a}cher, M.~Goebel, J.~Haller, A.~Hoecker, K.~M\"{o}nig and J.~Stelzer,
  ``Revisiting the Global Electroweak Fit of the Standard Model and Beyond with Gfitter,''
  \textit{Eur.\ Phys.\ J.\ C} {\bf 60}, 543 (2009)
  [Erratum: \textit{Eur.\ Phys.\ J.\ C} {\bf 71}, 1718 (2011)]
  [arXiv:0811.0009 [hep-ph]]. See also \url{http://project-gfitter.web.cern.ch}.
  
\bibitem{Aad:2012tfa} 
  G.~Aad {\it et al.} [ATLAS Collaboration],
  ``Observation of a new particle in the search for the Standard Model Higgs boson with the ATLAS detector at the LHC,''
  \textit{Phys.\ Lett.\ B} {\bf 716}, 1 (2012)
  [arXiv:1207.7214 [hep-ex]].

\bibitem{Chatrchyan:2012xdj} 
  S.~Chatrchyan {\it et al.} [CMS Collaboration],
  ``Observation of a new boson at a mass of 125 GeV with the CMS experiment at the LHC,''
  \textit{Phys.\ Lett.\ B} {\bf 716}, 30 (2012)
  [arXiv:1207.7235 [hep-ex]].
 \bibitem{ATLASH}ATLAS Higgs-boson publications \url{hhttp://j.mp/1PO6qgD} and other public results \url{http://j.mp/1A1kzA5}.

\bibitem{CMSH} CMS Higgs-boson papers \url{http://j.mp/1PO6I6V}.  
  
 
  \bibitem{Chatrchyan:2013iaa} 
  S.~Chatrchyan {\it et al.} [CMS Collaboration],
  ``Measurement of Higgs boson production and properties in the $WW$ decay channel with leptonic final states,''
  \textit{JHEP} {\bf 1401}, 096 (2014)
  [arXiv:1312.1129 [hep-ex]].
  

  
\bibitem{ATLAS:2014aga} 
  G.~Aad {\it et al.} [ATLAS Collaboration],
  ``Observation and measurement of Higgs boson decays to $WW^*$ with the ATLAS detector,''
  \textit{Phys.\ Rev.\ D}{\bf 92}, 012006 (2015)
  [arXiv:1412.2641 [hep-ex]].
   
 
\bibitem{Khachatryan:2014kca} 
  V.~Khachatryan {\it et al.} [CMS Collaboration],
  ``Constraints on the spin-parity and anomalous $HVV$ couplings of the Higgs boson in proton collisions at 7 and 8 TeV,''
  \textit{Phys.\ Rev.\ D}{\bf 92}, 012004 (2015)
  [arXiv:1411.3441 [hep-ex]].
\bibitem{Aad:2015mxa} 
  G.~Aad {\it et al.} [ATLAS Collaboration],
  ``Study of the spin and parity of the Higgs boson in diboson decays with the ATLAS detector,''
  \textit{Eur.\ Phys.\ J.\ C} {\bf 75}, 476 (2015)
  [arXiv:1506.05669 [hep-ex]].
  

  \bibitem{Aad:2015zhl} 
  G.~Aad {\it et al.} [ATLAS and CMS Collaborations],
  ``Combined Measurement of the Higgs Boson Mass in $pp$ Collisions at $\sqrt{s}=7\hbox{ and }8\tev$ with the ATLAS and CMS Experiments,''
  \textit{Phys.\ Rev.\ Lett.}  {\bf 114}, 191803 (2015)
  [arXiv:1503.07589 [hep-ex]].


 
\bibitem{mucombo} 
  The ATLAS and CMS Collaborations,
  ``Measurements of the Higgs boson production and decay rates and constraints on its couplings from a combined ATLAS and CMS analysis of the LHC $pp$ collision data at $\sqrt{s} = 7\hbox{ and 8}\tev$,''
  ATLAS-CONF-2015-044, 15 September 2015, \url{http://cds.cern.ch/record/2052552}.
 
\bibitem{Dittmaier:2012vm} 
  S.~Dittmaier {\it et al.},
  ``Handbook of LHC Higgs Cross Sections: 2. Differential Distributions,''
  arXiv:1201.3084 [hep-ph];
 ``SM Higgs Branching Ratios and Partial-Decay Widths (2012 update),'' \url{http://j.mp/1mirOhP}. 
 
\bibitem{Kauer:2012hd} 
  N.~Kauer and G.~Passarino,
  ``Inadequacy of zero-width approximation for a light Higgs boson signal,''
  \textit{JHEP} {\bf 1208}, 116 (2012)
  [arXiv:1206.4803 [hep-ph]].
 
\bibitem{Caola:2013yja} 
  F.~Caola and K.~Melnikov,
  ``Constraining the Higgs boson width with $ZZ$ production at the LHC,''
  \textit{Phys.\ Rev.\ D}{\bf 88}, 054024 (2013)
  [arXiv:1307.4935 [hep-ph]].
 
\bibitem{Khachatryan:2014iha} 
  V.~Khachatryan {\it et al.} [CMS Collaboration],
  ``Constraints on the Higgs boson width from off-shell production and decay to $Z$-boson pairs,''
  \textit{Phys.\ Lett.\ B} {\bf 736}, 64 (2014)
  [arXiv:1405.3455 [hep-ex]].
 
\bibitem{Aad:2015xua} 
  G.~Aad {\it et al.} [ATLAS Collaboration],
  ``Constraints on the off-shell Higgs boson signal strength in the high-mass $ZZ$ and $WW$ final states with the ATLAS detector,''
  \textit{Eur.\ Phys.\ J.\ C} {\bf 75}, 335 (2015)
  [arXiv:1503.01060 [hep-ex]].

\bibitem{Quigg:2009xr} 
  C.~Quigg and R.~Shrock,
  ``Gedanken Worlds without Higgs: QCD-Induced Electroweak Symmetry Breaking,''
 \textit{Phys.\ Rev.\ D}{\bf 79}, 096002 (2009)
  [arXiv:0901.3958 [hep-ph]].

\bibitem{Dawson:2013bba} 
  S.~Dawson {\it et al.},
  ``Higgs Working Group Report of the Snowmass 2013 Community Planning Study,''
  arXiv:1310.8361 [hep-ex].
  
\bibitem{ilc} The International Linear Collider, \url{https://www.linearcollider.org/ILC}

\bibitem{cepc} The Circular Electron--Positron Collider, \url{http://cepc.ihep.ac.cn}.

\bibitem{fcc-ee} The FCC-ee Design Study, \url{http://tlep.web.cern.ch/}.

\bibitem{Gourlay} S.~Gourlay, ``High Field Magnets for $pp$ Colliders,'' talk at the Hong Kong University of Science and Technology Jockey Club Institute for Advanced Study    Program on \textit{The future of high energy physics,} January 5-30
2015, \url{http://j.mp/1WBWiYR}.

\bibitem{Eichten:1984eu} 
  E.~Eichten, I.~Hinchliffe, K.~D.~Lane and C.~Quigg,
  ``Super Collider Physics,''
  \textit{Rev.\ Mod.\ Phys.} {\bf 56}, 579 (1984)
  [Erratum: \textit{Rev.\ Mod.\ Phys.}  {\bf 58}, 1065 (1986)].
  
\bibitem{HXSWG} LHC Higgs Cross Section Working Group, ``Higgs cross sections for HL-LHC and HE-LHC,'' \url{http://j.mp/1ZXfNM6}.
  
  \bibitem{FELIX} For example, FELIX: Proposal for a Forward ELastic and Inelastic EXperiment at the LHC, \url{http://felix.web.cern.ch}.
  
\bibitem{Arkani-Hamed:2015vfh} 
  N.~Arkani-Hamed, T.~Han, M.~Mangano and L.~T.~Wang,
  ``Physics Opportunities of a 100 TeV Proton-Proton Collider,''
  arXiv:1511.06495 [hep-ph].
 
\bibitem{Dine:2015xga} 
  M.~Dine,
  ``Naturalness under Stress,''
  \textit{Ann.\ Rev.\ Nucl.\ Part.\ Sci.}  {\bf 65}, 43 (2015)
  [arXiv:1501.01035 [hep-ph]].
   
  \bibitem{Quigg:2009vq} 
  C.~Quigg,
  ``Unanswered Questions in the Electroweak Theory,''
  \textit{Ann.\ Rev.\ Nucl.\ Part.\ Sci.}  {\bf 59}, 505 (2009)
  [arXiv:0905.3187 [hep-ph]];
  ``Particle Physics after the Higgs-Boson Discovery: Opportunities for the Large Hadron Collider,''
  \textit{Contemporary Physics} \textbf{57}, nnn (2106)
  arXiv:1507.02977 [hep-ph].
 
  \bibitem{wigner} E.~P.~Wigner, ``The Unreasonable Effectiveness of Mathematics in the Natural Sciences,'' \textit{Commun. Pure Appl. Math.} \textbf{13}, 1--14 (1960) \url{http://j.mp/1PngVae}.
 
 
  
\bibitem{Wilson:2004de} 
  K.~G.~Wilson,
  ``The Origins of Lattice Gauge Theory,''
  \textit{Nucl.\ Phys.\ Proc.\ Suppl.}  {\bf 140}, 3 (2005)
  [hep-lat/0412043].
  
\bibitem{Koppen}
  P.~Koppenburg, ``CP violation and CKM physics (including LHC$b$ news from run 2),'' LHCb-PROC-2015-029, \url{https://cds.cern.ch/record/2039659}.
  

\bibitem{Khachatryan:2014gha} 
  V.~Khachatryan {\it et al.} [CMS Collaboration],
  ``Search for massive resonances decaying into pairs of boosted bosons in semi-leptonic final states at $\sqrt{s} =8\tev$,''
  \textit{JHEP} {\bf 1408}, 174 (2014)
  [arXiv:1405.3447 [hep-ex]].

  \bibitem{Aad:2015owa} 
  G.~Aad {\it et al.} [ATLAS Collaboration],
  ``Search for high-mass diboson resonances with boson-tagged jets in proton-proton collisions at $ \sqrt{s}=8 \tev$ with the ATLAS detector,''
  \textit{JHEP} {\bf 1512}, 055 (2015)
  [arXiv:1506.00962 [hep-ex]].

\bibitem{Aaij:2015bfa} 
  R.~Aaij {\it et al.} [LHCb Collaboration],
  ``Determination of the quark coupling strength $|V_{ub}|$ using baryonic decays,''
  \textit{Nature Phys.}  {\bf 11}, 743 (2015)
  [arXiv:1504.01568 [hep-ex]].
  
\bibitem{Aaij:2014ora} 
  R.~Aaij {\it et al.} [LHCb Collaboration],
  ``Test of lepton universality using $B^{+}\rightarrow K^{+}\ell^{+}\ell^{-}$ decays,''
  \textit{Phys.\ Rev.\ Lett.}  {\bf 113}, 151601 (2014)
  [arXiv:1406.6482 [hep-ex]].
  
  \bibitem{ATLASdigamma} 
  The ATLAS collaboration,
  ``Search for resonances decaying to photon pairs in $3.2\fb^{-1}$ of $pp$ collisions at $\sqrt{s} = 13\tev$ with the ATLAS detector,''
  ATLAS-CONF-2015-081, \url{http://cds.cern.ch/record/2114853}.
  
\bibitem{CMS:2015dxe} 
  CMS Collaboration [CMS Collaboration],
  ``Search for new physics in high mass diphoton events in proton-proton
  collisions at $13\tev$,''   CMS-PAS-EXO-15-004, \url{http://cds.cern.ch/record/2114808}.
     
   \bibitem{Hinchliffe:2015qma} 
  I.~Hinchliffe, A.~Kotwal, M.~L.~Mangano, C.~Quigg and L.~T.~Wang,
  ``Luminosity goals for a 100-TeV $pp$ collider,''
 \textit{ Int.\ J.\ Mod.\ Phys.\ A}{\bf 30}, 1544002 (2015)
  [arXiv:1504.06108 [hep-ph]].
 
  \bibitem{Nadolsky:2008zw} 
  P.~M.~Nadolsky, H.~L.~Lai, Q.~H.~Cao, J.~Huston, J.~Pumplin, D.~Stump, W.~K.~Tung and C.-P.~Yuan,
  ``Implications of CTEQ global analysis for collider observables,''
  \textit{Phys.\ Rev.\ D}{\bf 78}, 013004 (2008)
  [arXiv:0802.0007 [hep-ph]].
  
  \bibitem{Tye:2015tva} 
  S.-H.~H.~Tye and S.~S.~C.~Wong,
  ``Bloch Wave Function for the Periodic Sphaleron Potential and Unsuppressed Baryon and Lepton Number Violating Processes,''
  \textit{Phys.\ Rev.\ D}{\bf 92}, 045005 (2015)
  [arXiv:1505.03690 [hep-th]].

 
\bibitem{Campbell:2011bn} 
  J.~M.~Campbell, R.~K.~Ellis and C.~Williams,
  ``Vector boson pair production at the LHC,''
  \textit{JHEP} {\bf 1107}, 018 (2011)
  [arXiv:1105.0020 [hep-ph]]. See also \url{http://mcfm.fnal.gov}.
 
  \bibitem{benedikt}
M. Benedikt, ``FCC study overview and status,'' talk at FCC Week
2015, Washington D.C., 23--29 March 2015, 
\url{http://j.mp/1Upsuhp}. 

\bibitem{Raby:2009sf} 
  S.~Raby, M.~Ratz and K.~Schmidt-Hoberg,
  ``Precision gauge unification in the MSSM,''
  \textit{Phys.\ Lett.\ B} {\bf 687}, 342 (2010)
  [arXiv:0911.4249 [hep-ph]].
  
\bibitem{ATLAS:2013lla} 
  ATLAS Collaboration,
  ``Measurement of multi-jet cross-section ratios and determination of the strong coupling constant in proton-proton collisions at $\sqrt{s} = 7\tev$ with the ATLAS detector,''
  ATLAS-CONF-2013-041, \url{http://cds.cern.ch/record/1543225}.
  

\bibitem{Khachatryan:2014waa} 
  V.~Khachatryan {\it et al.} [CMS Collaboration],
  ``Constraints on parton distribution functions and extraction of the strong coupling constant from the inclusive jet cross section in $pp$ collisions at $\sqrt{s} = 7\tev$,''
  \textit{Eur.\ Phys.\ J.\ C} {\bf 75}, 288 (2015)
  [arXiv:1410.6765 [hep-ex]].

\end{thebibliography}
\end{document}